\documentclass[prb,onecolumn,epsf,aps,superscriptaddress,showpacs]{revtex4}
\usepackage{epsfig}
\usepackage{amsmath}
\usepackage{sidecap}
\usepackage[figuresleft]{rotating}
\usepackage{caption}
\usepackage[]{graphicx}
\usepackage{lscape}
\newcommand{\EssD}{\mathcal{D}}

\def\prb{Phys. Rev. B }

\def\prl{Phys. Rev. Lett. }

\begin{document}
\title {Controllable $\pi$ junction in a Josephson quantum-dot device with molecular spin.}
\author{C. Benjamin}
\email{cbiop@yahoo.com} \affiliation{Centre de Physique Th\'eorique,
Case 907, Luminy, 13288 Marseille Cedex 9, France}
 \author{T. Jonckheere}
\affiliation{Centre de Physique Th\'eorique,
Case 907, Luminy, 13288 Marseille Cedex 9, France}
\author{A. Zazunov}
\affiliation{Centre de Physique Th\'eorique,
Case 907, Luminy, 13288 Marseille Cedex 9, France}
\affiliation{LEPES, 25 Avenue des Martyrs, 38000 Grenoble, France}
\author{T. Martin}
\affiliation{Centre de Physique Th\'eorique,
Case 907, Luminy, 13288 Marseille Cedex 9, France}
\affiliation{Universit\'e de la M\'edit\'erann\'ee, 13288 Marseille cedex 9, France}

\date{\today}

 \begin{abstract}
We consider a model for a single molecule with a large frozen spin
sandwiched in between two BCS superconductors at equilibrium, and
show that this system has a  $\pi$ junction behavior at low
temperature. The $\pi$ shift can be reversed by varying the other
parameters of the system, e.g., temperature or the position of the
quantum dot level, implying a controllable $\pi$ junction with
novel application as a Josephson current switch. We show that the mechanism 
leading to the $\pi$ shift can be explained simply in terms of the contributions 
of the Andreev bound states and of the continuum of states above the superconducting gap.
The free energy for certain configuration of parameters shows a
bistable nature, which is a necessary pre-condition for
achievement of a qubit.
\end{abstract}
\maketitle
\section{Introduction}

 Molecular spintronics is a promising new domain, at the
convergence of two challenging disciplines. On the one hand
there is molecular electronics, where single molecules are used to create
electronics devices at the nanometric scale with unique
properties, while on the other hand there is spintronics, where the spin of
the electron is used as the relevant quantity in place of the
electronic charge. The latter allows us to take advantage of the
unusual properties of spin, like a long coherence time. It is in
this context that we consider in our work the equilibrium
properties of a molecule with a large magnetic moment placed
between two superconductors, when a Josephson current flows
between the two superconductors through the molecule. The focus
will be on the effect, on the Josephson current, of the coupling
between the spin of the electrons producing the current and the
molecular spin. We will show in particular that when this spin
coupling is large enough, the superconducting junction behaves as
a $\pi$ junction, with a reversal of the Josephson current
compared to the case without spin coupling.

Of great importance for molecular spintronics are molecules
possessing a large spin, or ``single molecule magnets''. Such
molecules can now be synthesized, for example the molecule Mn12ac,
which has a ground state with a large spin $S=10$, and a very slow
relaxation of magnetization at low temperature.\cite{gatteschi}
This slow relaxation is due to a high anisotropy barrier for the
magnetization, around $5.6$ meV for Mn12ac. For the system we are
considering, this is a very large energy, as the typical energies
in our system (temperature, coupling to the electrodes, etc.) are
at most of the order of the superconducting gap (which is $0.1$
meV in Aluminum for example). This motivates our choice to take
the molecular spin as a fixed quantity, which will act as a local
magnetic field for the electrons going through the molecule.
Note that other systems, where the spin is not fixed, 
involving for example superconducting transport through 
fullerene molecule doped with magnetic impurities
have been considered experimentally and theoretically.\cite{bouchiat05,levy06}
Concerning the electronic transport across the molecule, we model
the molecule as a single resonant level, i.e. a quantum dot. As we
will be interested in the regime of good transparency between the
molecules and the superconducting electrodes, we will neglect in
this work the electronic interactions on the resonant
level.\cite{veciono}

The main result of our paper is to show that, when the coupling to
the molecular spin is large enough, the system shows a $\pi$
shift. A reversal of the super-current in a Josephson device and
the free energy having a global minima at phase difference $\pi$
is referred to as $\pi$ shift and a Josephson junction displaying
this is termed a $\pi$ junction.\cite{golubov} The $\pi$ junction
has potential applications in superconducting electronics, in
quantum logic circuits as switches and are an integral part of
superconducting phase qubits. We will also show that this $\pi$
shift can be controlled by the other parameters of the system
(position of the dot level, temperature, coupling to the
electrodes, etc.), allowing to reverse the $\pi$ shift and recover
a standard Josephson junction.

Generally in works related to $\pi$ junction behavior  the bound
state current, which is due to current carrying Andreev bound
states formed between the two superconductors, is investigated
while the contribution from the continuum of states above the
superconducting gap is ignored.\cite{beena}
 There are good reasons for doing so, since the continuum
contribution is generally much reduced compared to bound state
current, especially in the limit of a long junction or a very
short one.  However recent works have shown that the continuum
current cannot be ignored,\cite{heikkila,kamenev,kirch,jens} especially in
a Josephson junction which is neither very short nor very long. In
this work, we calculate explicitly the contribution from the
continuum, and we show that the in the presence of a large
coupling to the molecular spin, the continuum current is essential
to understand the $\pi$ junction behavior. In some regime, the
bound state current can even vanish, and the continuum current is
then the only contribution. We will show also that, with some
fine-tuning of the parameters of the system, the system can be in
a bi-stable state, where the $\phi=0$ and $\phi=\pi$ state are
equally stable; this bi-stability is a necessary condition for a
possible qubit implementation.

\begin{figure}[!]
\protect\centerline{\epsfxsize=4.0in\epsfbox{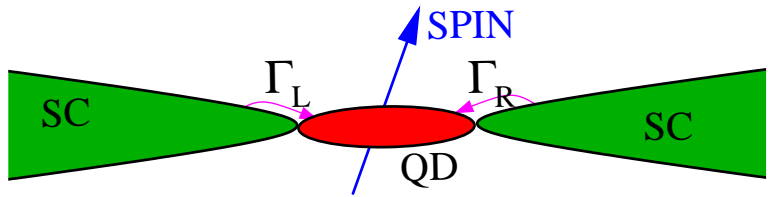}}
\caption{Our model system}
\end{figure}

The rest of the work is organized as follows. The next section
deals with a short history of the $\pi$-shift as seen in Josephson
junctions and with the possible applications which such behavior
may have. Section III is devoted to the derivation of the
Josephson current when coupling to the molecular spin is present.
In section IV we use the formulae obtained in section III to show
the behavior of the Josephson current as a function of the
coupling to the molecular spin, and of the other parameters. We
give a detailed explanation of the mechanism leading to the $\pi$
shift. In section V we discuss some potential applications of our
system, first as a Josephson current switch, then as a
superconducting qubit. Section VI is devoted to concluding
remarks.

\section{Brief history of $\pi$-shift}

In order to show how our work and results differ from existing
works on $\pi$ junctions, we find useful to give a very short
history of the $\pi$-shift. $\pi$ junctions were first proposed
theoretically by Bulaevskii and coworkers in
Ref.[\onlinecite{Bulaevskii}]. They considered a tunnel junction
with magnetic impurities in the barrier. In this system spin-flip
tunneling leads to a formation of $\pi$ junction. They also
predicted that a super-conducting ring containing a $\pi$ junction
could generate a spontaneous current and a magnetic flux opening
the way for experimental detection. Spin flip tunneling in
superconductor-quantum dot-superconductor(S-QD-S) system has also
been shown to give rise to a $\pi$ junction behavior as in
Refs.[\onlinecite{glaz,siano,pan}]. It was Kulik who in 1966 was
the first to discuss the spin-flip tunneling through an insulator
with magnetic impurities.\cite{kulik} The spin-flip tunneling is
predicted to dominate the Josephson current when spin on the
quantum dot is non-zero. In S-QD-S junction, changes in the sign
of the critical current could be observed as a function of the
quantum dot gate voltage which controls the occupancy of a quantum
dot. Due to this gating capability one has more control over the
magnetic state of a barrier in S-QD-S junction compared to a
magnetically doped Superconductor-Insulator-Superconductor
junction.\cite{arovas}
Superconductor-Ferromagnet-Superconductor(SFS) have also been
shown to give rise to a $\pi$ junction behavior both in
theory\cite{buzdin,sell,radovic} as  well as in
experiments.\cite{ryazanov,frolov}  The study of the
superconducting $\pi$ state sheds more light on the coexistence of
superconductivity and ferromagnetism in general and is also
important for superconducting electronics.\cite{buz-baladie}
Generally with increase in the strength of the exchange field the
$\pi$ shift is observed with a reversal and suppression of the
super-current. In SFSFS systems with the ferromagnets in
anti-parallel alignment, however, with increase in the strength of
the exchange field an increase in super-current is
observed.\cite{bergeret} In SIFIS and SFcFS, where c denotes a
constriction, structures also such $\pi$ junctions have been
observed.\cite{golubov} Recently triplet
superconductor-ferromagnet-triplet superconductor junctions have
been predicted to have potential applications as current
switches.\cite{boris} In contrast to SFS systems, $\pi$ junction
behavior in SNS systems occurs due to the creation of a
non-equilibrium distribution of electrons in the barrier via a
control channel.\cite{baselmans}  Thus in these systems the
$\pi$-junction can be controlled via a voltage applied to the
control channel this makes such devices ideal for them to be used
in superconducting digital circuits, especially as a phase
inverter, i. e., $\pi$-SQUID\cite{guichard} for complementary
Josephson digital devices. Further application of $\pi$ junctions
as candidates for engineering quantum bits have been
predicted.\cite{ioffe} Finally, $\pi$ junctions have also been
theoretically predicted and experimentally observed in
superconducting d-wave junctions.\cite{golubov}

\section{Derivation of the Josephson current}

\subsection{Model Hamiltonian}

The Josephson Current($I_J$) can be calculated from the derivative of the
free energy(F) with respect to the phase difference($\phi$) across
the superconducting leads $I_{J}=2\frac {dF}{d\phi}$, in
equilibrium. The free energy in turn is defined as $F=-kT \ln Z$,
where Z is the partition function of our system. Thus calculating
Z is the first step in calculating the Josephson current in our
system. The full Hamiltonian of our system is written as below:

\begin{equation}
H=H_{D-S}+\sum_{j=L,R} H_{j}+H_{T},
\end{equation}
where $H_{D-S}$ defines the Hamiltonian of the quantum-dot
molecule with spin, $H_{j}$ represents the superconducting leads,
while $H_T$ denotes the tunneling part. The dot-spin Hamiltonian
is:
\begin{equation}
 H_{D-S}=\epsilon \sum_{\sigma}
d_{\sigma}^{\dagger}d_{\sigma}+J S (d_{\uparrow}^{\dagger}d_{\uparrow}- d_{\downarrow}^{\dagger}d_{\downarrow})
\end{equation}
where $d_{\sigma}, d_{\sigma}^{\dagger}$ are the electronic
operators in the dot,  $\epsilon$ is the energy level of the dot,
$J$ the coupling between the molecular spin ${\bf S}$ and the
electronic spin on the dot level. The coupling term comes from the
exchange interaction $J {\bf S} \cdot {\bf s}$, where ${\bf s}$ is
the electronic spin on the dot, but as explained in the
introduction, the molecular spin is fixed in our system, and we
chose the spin quantization axis  along the spin orientation.
 In the superconducting Hamiltonian it is convenient to perform a gauge transformation
which removes the phase from the order parameter.\cite{schrieffer}
Thus,
\begin{equation} H_{j}=\sum_{k}
\Psi_{jk}^{\dagger}(\xi_{k}\sigma_{z}+\Delta \sigma_{x})
\Psi_{jk}, \Psi_{jk}=\left(\begin{array}{c} \psi_{jk,
\uparrow}\\
\psi_{j(-k),\downarrow}^{\dagger}
\end{array} \right)
\label{eq:Sspinor}
\end{equation}
and finally the tunnelling part can be written in the standard
form with a hopping parameter $t_j$ determining the transfer
properties of the junction. The Pauli matrices mentioned in the
above equation are matrices in particle-hole space. The effect of
the gauge transformation on the tunnel Hamiltonian is the
appearance of a phase dependence in the hopping parameter
\begin{equation}
 H_{T}=\sum_{jk}\Psi_{j k}^{\dagger} T_{j} d + h.c.,
\quad \quad  d_{}=\left(\begin{array}{c} d_{\uparrow}\\
d_{\downarrow}^{\dagger}
\end{array} \right)
\label{eq:dspinor}
\end{equation}
with $ T_{L}=t_{L}\sigma_{z}e^{i\sigma_{z}\phi/4},
T_{R}=t_{R}\sigma_{z}e^{-i\sigma_{z}\phi/4}$, where $\phi$ is the
phase difference between the superconducting leads, and $t_j$ is
the tunnelling amplitude between the $j$th lead and dot.

\subsection{Neglecting Coulomb interaction}

In the present work we chose to neglect the charging energy of the
molecule. Molecular electronics transport calculations are
typically concerned with two limits: either the limit where the
tunnelling rate from the molecule to the leads dominates over the
charging energy, or the opposite limit of strong Coulomb blockade
which can be dealt in the incoherent regime or in the coherent
regime. The validity of each regime depends on the transparency of
the tunnel barriers connected to the molecule, and on the
capacitances seen by the molecular quantum dot with respect to the
leads. So far in the literature, many efforts have focused on the
Landauer-Buttiker description of molecular electronics transport.
\cite{landauer_buttiker} In some instances\cite{fagas} this
approach is supplemented by taking Coulomb interactions in an
effective manner using density functional theory, but the Green
function which is used to compute the transmission coefficient is
determined from a specific electron configuration, as for an
effective one-electron Coulomb potential. Here we adopt a
Hamiltonian approach to compute the current, but the basic
assumptions for neglecting Coulomb blockade effects also apply.
This is justified as follows.

In fact this stems from the fact that the charging energy of the
molecular quantum dot is assumed to be small compared to the the
escape rate of the electrons from the dots to the leads, later to
be referred to as $\Gamma$. This regime triggers a substantial
broadening of the dot level, as for instance was discussed in
Ref.[\onlinecite{guyon_mujica}] where a molecule was attached to
metallic substrate, being effectively ``metalized''. Qualitatively
speaking, if the transparency from the dot to the leads is close
to unity, the time scale which characterizes the lifetime of
electrons on the molecular quantum dot is so short that Coulomb
blockade effects do not have time to operate.

These qualitative arguments are substantiated by theoretical works
on the Coulomb blockade in the presence of highly transmissive
barriers, which were carried out more than a decade ago.
 Ref.[\onlinecite{flensberg}] studies the behavior of a quantum dot
embedded between two point contacts, and finds a crossover to a
regime where charge fluctuations in the dot are dominant,
therefore wiping out charge quantization (Coulomb blockade)
effects. The temperature which characterizes this crossover is of
course proportional to the dot charging energy, but it also goes
to zero in the limit of ideal transmission. Ref.[\onlinecite{matveev}]
 considers a quantum dot connected to single
barriers, and shows that the energy of the dot undergoes (Coulomb
blockade) oscillations as a function of gate voltage as long as
the transmission coefficient of the barrier, which isolates the
dot, does not approach unity. On another note, Ref.[\onlinecite{nazarov}]
 uses a path integral framework\cite{schon_zaikin} to describe a quantum dot with
arbitrary barriers. Previous results\cite{flensberg} concerning
charge fluctuations are recovered, but more importantly it is
shown that the effective charging energy is exponentially reduced
at ideal transmission. This latter result applies, granted to a
dot coupled to several channels, but these features are expected
to survive for a a dot coupled to a single, highly transmissive
channel.

Note that in the above, it is sufficient for only one of the
two contacts to have a large capacitance in order to be able to
neglect Coulomb effects. The present point of view is consistent with recent works on molecular
electronics issues where phonons are involved \cite{other_no_coulomb},
but electron-electron interactions are neglected nevertheless.

Finally, we stress the fact that there exist actual experiments
in molecular electronics, which can achieve the high transmission
conditions which are assumed in the present work.
It has been demonstrated that break junction geometries\cite{ruitenbeek}
can achieve close to ideal transmission: for a Hydrogen molecule
which is sandwiched between Platinum electrodes, one observes a conductance
quantization plateau at $2e^2/h$, when bringing the two electrodes together,
as expected from a monovalent metal as Hydrogen (in contrast, a pure platinum
junction yields steps at $4e^2/h$). This is the direct evidence
of single, perfectly transmitting channel.
To summarize, in these works, which have been extended to study the effects
of vibrations on the molecule\cite{ruintenbeek2}, no Coulomb
blockade effects show up at all.
For the present study, we thus assume that our molecule is placed
under the same conditions as in these break junction experiments.
Note that high transmission conditions have also been obtained with Carbon nanotubes\cite{bouchiat99,bouchiat01}.

\subsection{Effective action}

To calculate the partition function we use the path integral
approach. In this method the partition function is given by:
\begin{equation}
Z=\int \prod_{jk} (\EssD {\bar\Psi_{jk}} \EssD {\Psi_{jk}}\EssD
{\bar d}\; \EssD {d})\; e^{-S_A}
\end{equation}

Z is written as a functional integral over grassmann fields for
the electronic degrees of freedom ($\Psi, \bar\Psi $). The
Euclidean action reads:
\[S_A=S_{D}+\int_{0}^{\beta} \! d \tau
[\sum_{jk}{\bar\Psi_{jk}(\tau)}(\partial_{\tau}+\xi_{k}\sigma_{z}+\Delta\sigma_{x})\Psi_{jk}(\tau)+{
H_{T}(\tau)}] \] $\beta$ is the inverse temperature, and
${H_{T}(\tau)}=\sum_{jk}{\bar\Psi_{jk}(\tau)}T_{j}d(\tau)+h.c.
 \mbox{ while } S_{D}=\int_{0}^{\beta}\! d\tau [{\bar d}(\partial_{\tau}+\epsilon
\sigma_{z}+J S)d]$.
 After integrating out the leads we get
\begin{equation}
Z=\int\! \EssD {\bar d}\; \EssD d \; e^{-S_{eff}} \quad \mbox{with} \quad \quad
S_{eff}=S_{D}-\int_{0}^{\beta}\! d\tau \; d\tau' \;{\bar d(\tau)}{\check
\Sigma(\tau - \tau ')} d(\tau ')
\end{equation}
where $\check
\Sigma(\tau)=\sum_{j=L,R} T_{j}^{\dagger} G(\tau) T_{j}$ and
$G(\tau)=\sum_{k}
(\partial_{\tau}+\xi_{k}\sigma_{z}+\Delta\sigma_{x})^{-1}
\delta(\tau)$.

We perform a Fourier transform on the Matsubara frequencies (with  $w_{n}=(2n+1)\pi/\beta$):
$\delta(\tau)=\frac{1}{\beta} \sum_{w_n} e^{-i w_{n} \tau}$ and
$G(\tau)=\frac{1}{\beta} \sum_{w_n} e^{-i w_{n} \tau} G_{w_n}$, which gives for the Green function $G$:
\begin{eqnarray}
G_{w_n}&=&\int\! d \xi\; \nu(\xi)(-i
w_{n}+\xi_{k}\sigma_{z}+\Delta\sigma_{x})^{-1}
\simeq \frac{\pi
\nu(0)}{\sqrt{w_{n}^{2}+\Delta^{2}}}(iw_{n}+\Delta \sigma_{x})
\end{eqnarray}
In the above equation, $\nu(\xi)=\sum_{k}\delta(\xi-\xi_{k})$ is
approximated as a constant $\nu(0)$, the density of states at the
Fermi level in the normal leads. This gives for the self-energy:
\begin{equation}
{\check \Sigma_{w_n}}=\frac{{
\Gamma/2}}{\sqrt{\Delta^{2}-(iw)^{2}}}[iw_{n}-\Delta \cos (\phi/2)
\sigma_{x} -\gamma \Delta \sin (\phi/2) \sigma_{y}]
\end{equation}
with $\gamma=\frac {\Gamma_{L}-\Gamma_{R}}{\Gamma_{L}+\Gamma_{R}},
{ \Gamma}=\Gamma_{L}+\Gamma_{R}, \Gamma_{L/R}=2\pi \nu(0)
t_{L/R}^{2}$. We get finally for the effective action
(introducing $d(\tau)=\frac{1}{\sqrt{\beta}}\sum_{w_{n}} e^{-i w_{n} \tau} d_{w_{n}}$)
\begin{equation}
\label{eq:action_matsu}
S_{eff}=\sum_{w_{n}} \bar{d}_{w_{n}} {\cal M}_{w_n} d_{w_{n}} \quad \mbox{with}\quad
{\cal M}_{w_n} = -i w_{n}+\epsilon
\sigma_{z}+J S-{\check \Sigma_{w_{n}}}\end{equation}

\subsection{ Andreev Levels}
The dispersion equation for the Andreev levels is given by the
eigenvalues of the effective action in Eq.~(\ref{eq:action_matsu})
(with $iw=z$)
\begin{equation}
\label{eq:M} \det \left[z-\epsilon\sigma_{z} -J S+\frac{
\Gamma/2}{\sqrt{\Delta^{2}-z^{2}}}(z-\Delta\cos(\phi/2)\sigma_{x}
-\gamma\Delta\sin(\phi/2)\sigma_{y})\right]=0
\end{equation}
which gives (introducing the parameter $s= J S$):
\begin{equation}
\label{eq:detM}
(z+\frac{\Gamma
z}{2\sqrt{\Delta^{2}-z^{2}}}
-\epsilon-s)(z+\frac{\Gamma
z}{2\sqrt{\Delta^{2}-z^{2}}}
+\epsilon-s)-\frac{\Gamma^{2}\Delta^{2} (\cos^{2}(\phi/2)+\gamma^{2}\sin^{2}(\phi/2))}{4(\Delta^{2}-z^2)} =0
\end{equation}
While this cannot be solved analytically in general, there are two limiting regimes where one can get
an analytical expression of the solutions, giving two Andreev levels. For simplicity, we choose here $\epsilon=0$ and $\gamma=0$.\\
case 1: ${ \Gamma}\gg \Delta:$
\begin{equation}
\label{eq:Elimit1}
z=E_{1,2} = \Delta \, \cos \left[ \mbox{Arccos}\left(\frac{\pm \cos {\phi/2}}{\sqrt{1 + 4 s^2 / \Gamma^2}}\right) +
  \mbox{Arctan}\left(\frac{2 s}{\Gamma} \right) \right]
\end{equation}
case 2: ${ \Gamma}, s \ll \Delta:$
\begin{equation}
\label{eq:Elimit2}
z=E_{1,2}=s \pm \frac{\Gamma}{2} \cos(\phi/2)
\end{equation}

In the general case, for arbitrary $\Gamma, \epsilon, s(=J S)$ and
$\Delta$, we calculate numerically the roots, by transforming the
l.h.s. of Eq.~(\ref{eq:detM}) into a 8th order polynomial in z to
get rid of the square roots, and then calculating the roots of
this polynomial. We find that only two of these roots correspond
to roots of Eq.~(\ref{eq:detM}) (see also ref.
[\onlinecite{kirch}]), and that these two roots are real and
belong to $[-\Delta,\Delta]$. There are thus always two Andreev
bound states, as in the zero spin case: the effect of the spin
term is merely to move these two states, but it does not introduce
new bound states.

\begin{figure}[!]
\protect\centerline{\epsfxsize=4.0in\epsfbox{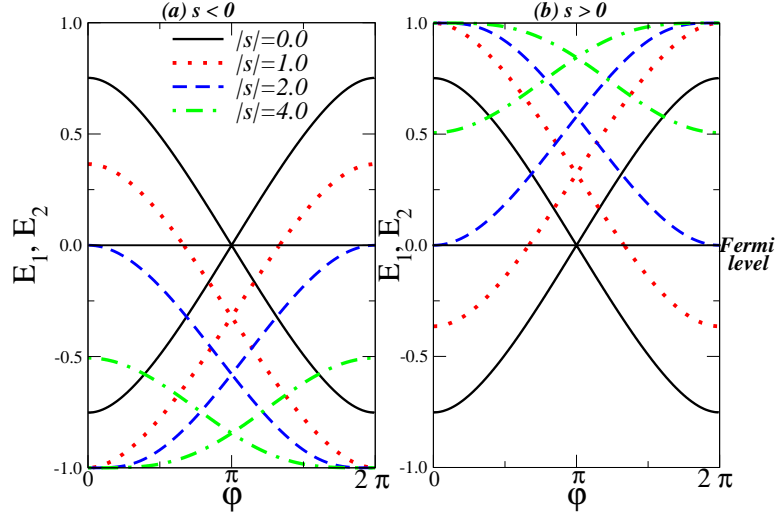}}
\caption{ The two Andreev bound states as function of phase
difference. For $s>0$ in the right panel and
for $s<0$ in the left panel.  The labelling
of the curves is as follows: spin s=0(black,solid line), 1.0
(red,dotted line), 2.0 (blue,dashed line) and 4.0 (green,
dot-dashed line). The other parameters are $\Gamma=4.0$,
$\gamma=0.0$, $\beta=1000$, $\epsilon=0.0$, and $\Delta=1.0$.}
\label{fig:Andreevlevelsphi}
\end{figure}

 In Fig.~\ref{fig:Andreevlevelsphi}, we plot the two Andreev bound
state positions as a function of the phase difference for four
values of spin, $|s|= 0, 1, 2, $ and $4$, for large transparency of
the contacts ($\Gamma=4.0$) and very low temperature
($\beta=1000$).
 $\Delta=1$ is taken as the unit of energy in our system, as in the rest of this work.
The right panel in Fig.~\ref{fig:Andreevlevelsphi} corresponds to
$s > 0$ while left panel is for $s < 0 $. We see that when the absolute value of $s$
is increased, the two Andreev levels are pushed towards $+\Delta$ or $- \Delta$.

It might seem surprising that the Andreev bound states are different for $+s$ and $-s$.
Indeed, our physical system is invariant under the interchange of spin up and spin down electron
combined to the exchange $s \to -s$; as the superconductors are the same under the spin interchange, our
physical system has thus to be invariant under the transformation $s \to -s$. As will be shown in the
next sections, the total Josephson current, which is a physically measurable quantity, is invariant under
this transformation ($s \to -s$). However, the expression of this total current in terms of the Andreev
bound state current and of the continuum current, depends on ths sign of $s$ (and so do the Andreev bound states).~\cite{jens}

The different Andreev bound states obtained for $s$ and for $-s$ are thus two different ways to represent the same physical situation.
The fact that we obtain one of the two possibilities for a given external spin can be traced back to our initial choice of the spinors
for the superconductors (Eq.~(\ref{eq:Sspinor})) and for the dot (Eq.~(\ref{eq:dspinor})): had we chosen the spinors defined with opposite
spin (for example, $d^{\dagger} = (d^{\dagger}_{\downarrow},d_{\uparrow})$, to be compared with the definition of Eq.~(\ref{eq:dspinor})),
 we would have obtained for $s<0$ the Andreev bound states
shown here for $s>0$, and vice-versa. One could also use a combination of the two possibilities of spinors, in order to get a spectrum
of Andreev bound states (and bound state current) which is independent of the sign of $s$; in this case, the spectrum is composed
of four Andreev bound states, which are precisely the two we obtained for $s>0$ plus the two for $s<0$.
In this paper, we have chosen to keep the spinors as defined in Eq.~(\ref{eq:Sspinor}) and  Eq.~(\ref{eq:dspinor}).
This choice will give us a particularly simple picture for the mechanism leading to the pi-shift (see below).

\subsection{Josephson Current}

 The partition function after integrating out the
$\{d,{\bar d}\}$ variables is given by-
\begin{equation}
Z=\int \!\EssD {\bar d}\;\EssD d\; e^{-S_{eff}}=\prod_{w_n} \det
{\cal M}_{w_{n}}
\end{equation}
where ${\cal M}_{w_n}$ is given in Eq.~(\ref{eq:action_matsu}).
The Josephson current then reduces to:
\begin{eqnarray}
\label{eq:IJmatsu}
I_{J}&=&-\frac{2}{\beta}\frac{\partial}{\partial \phi}  \ln Z=
-\frac{2}{\beta}\frac{\partial}{\partial \phi} \sum_{w_n} \ln
(\det {\cal M}_{w_n})\nonumber\\
&=&-\frac{1}{\beta} \sum_{w_n}
\frac{\Gamma^{2}\Delta^{2}(1-\gamma^{2})\sin(\phi)}{4[\det {\cal
M}_{w_n}](\Delta^{2}-(iw_{n})^{2})}=-\frac{1}{\beta}\sum_{w_{n}=-\infty}^{\infty}
f(iw_{n}) \end{eqnarray}
where the last equality defines the function $f$.

Further, the Free energy is given by:
\begin{equation}
\label{free_matsu}
 F=-\frac{1}{\beta}\sum_{w_n}\ln(\det {\cal M}_{w_n}).
\end{equation}
In the above equations, $\det {\cal M}_{w_n}$ is the same as the
l.h.s of Eq.~(\ref{eq:detM}), with $i w_n$ replacing $z$.

>From the above equation, one can calculate the total Josephson
current by summing over the Matsubara frequencies. However, we can
transform the above equation in order to separate explicitly the
contributions of the Andreev bound states and of the continuum,
which are physically meaningful. In order to calculate these
contributions, we take advantage of the fact that the Matsubara
frequencies are the poles of the Fermi function
$n_{f}(z)$~[\onlinecite{mahan}]. We then consider the integral
$I=\int_{C} \frac{dz}{2\pi i} f(z) n_{f}(z)$, where the function
$f(z)$ is defined in Eq.~(\ref{eq:IJmatsu}). The function $f(z)$
as seen earlier has two poles on the real axis between $-\Delta$
and $\Delta$ (these are simply the two Andreev bound states, for
which $\det \mathcal{M}=0$).
 Further, because of the square roots terms in the $\det
\mathcal{M}$, it has branch points at $z=\pm \Delta$; we have
chosen to place branch cuts on the real axis, for $z \in
[\Delta,\infty]$ and $z \in [-\infty,-\Delta]$. We thus chose the
contour C as two large semi-circles plus parts going around the
branch cuts. We illustrate the contour, poles and branch cuts in
Fig. 3. Thus integral I can be broken into the sum of the
contributions from the large circle $D$ of radius $R$, the two
small circles at $\pm \Delta$, denoted by $d_{1,2}$ of radius
$\epsilon$, and the contribution from the branch cuts. Therefore,
\begin{eqnarray}
 I=\frac{1}{2\pi i}\int_{C}f(z)n_{f}(z)dz&=&\frac{1}{2\pi i}\int_{D}\!\!dz \; f(z)n_{f}(z)+\frac{1}{2\pi i}
\int_{d_{1}}\!\!dz \; f(z)n_{f}(z) +\frac{1}{2\pi i}\int_{d_{2}}\!\!dz \; f(z)n_{f}(z)   \nonumber\\
&+& \frac{1}{2\pi i}\lim_{\epsilon\rightarrow
0}\int_{\Delta}^{\infty} \! \!\!\! dz \; [f(z+i\epsilon)- f(z-i\epsilon)]n_{f}(z)\nonumber\\
&+&\frac{1}{2\pi i}\lim_{\epsilon\rightarrow0}
\int_{-\infty}^{-\Delta} \!\!\!\!dz \; [f(z+i\epsilon)- f(z-i\epsilon)]n_{f}(z)
\end{eqnarray}
The integrals over $D$ and $d_{1,2}$ tend to zero as
$R\rightarrow \infty$ and $\epsilon \rightarrow 0$. The last two
terms in the above equation define the contribution from the
continuum to the current, which we denote as $I_c$.
\begin{figure}[!]
\protect\centerline{\epsfxsize=4.5in\epsfbox{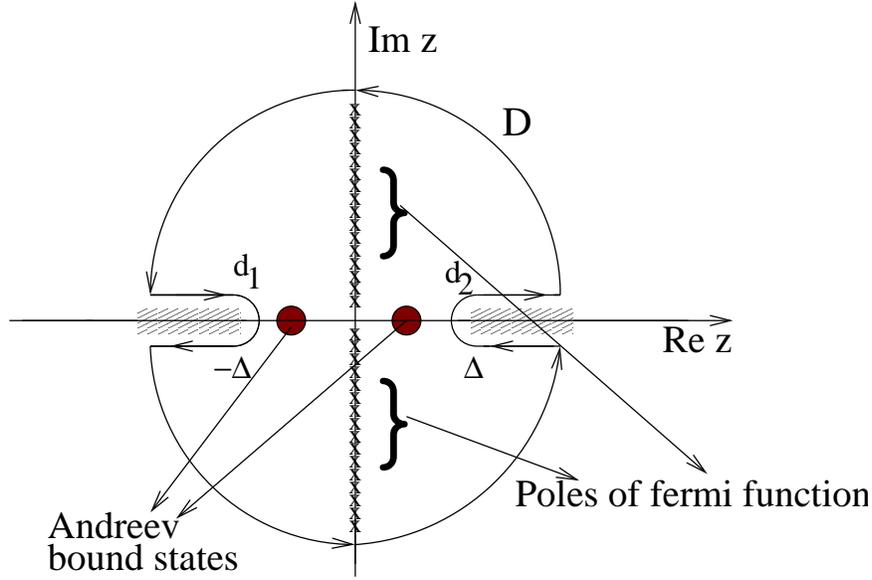}}
\caption{ The contour includes poles from the Fermi function, two
Andreev bound states and the contribution from the branch cuts. }
\end{figure}
 From Cauchy's residue theorem the integral I can also be evaluated
as
follows:\\
(a) Function $n_{f}(z)=\frac{1}{e^{\beta z}+1}$ has poles at
$z=i w_n$, with residue $-1/\beta$.
The contributions from these poles is thus:
$-\frac{1}{\beta}\sum_{w_{n}} f(iw_{n})$, which is precisely $I_J$ (Eq.~(\ref{eq:IJmatsu})).\\
(b) $f(z)$ has 2 poles of its own, written $E_1$ and $E_2$. These
gives the Andreev bound states contribution, which we denote by
$I_b$.
\\We have thus
\begin{equation}
\label{eq:IJ}
\left\{
\begin{split}
I_{J}& =I_{b}+I_{c}\\
I_{b}&=-n_F(E_1) \mbox{res}(f(E_1)) - n_F(E_2) \mbox{res}(f(E_2)) \\
I_{c}&=\lim_{\epsilon\rightarrow 0}\frac{1}{\pi}
\left[\int_{\Delta}^{\infty} \! \! dz + \int_{-\infty}^{-\Delta} \!\! dz \right]
\Im(f(z+i\epsilon)n_{f}(z+i\epsilon))
\end{split}
\right.
\end{equation}
where res denotes the residue of the quantity in square brackets,
and $\Im$ stands for the imaginary part. We see from the equation
giving $I_b$ that the contribution from each Andreev bound state
is simply proportional to the occupation number $n_F(E_i)$
$(i=1,2)$ of this level. Equation~(\ref{eq:IJ}) is the central
result of this work, which we have used to calculate numerically
the $I_b$ and $I_c$ curves shown in the following.

\section{Results}
In this section, we show the results obtained numerically for the
Josephson current using equation~(\ref{eq:IJ}). We will see that,
when the coupling to the spin is strong enough, the junction
behaves as a $\pi$ junction, and that the other parameters of the
system allow a control of this $\pi$ junction. We will also show
that the mechanism leading to this $\pi$ junction behavior can be
understood simply in terms of the bound states and continuum
contributions to the Josephson current.

\subsection{Total Josephson current and free energy: The
$\pi$ shift} 
\begin{figure}[!]
\protect\centerline{\epsfxsize=7.5in\epsfbox{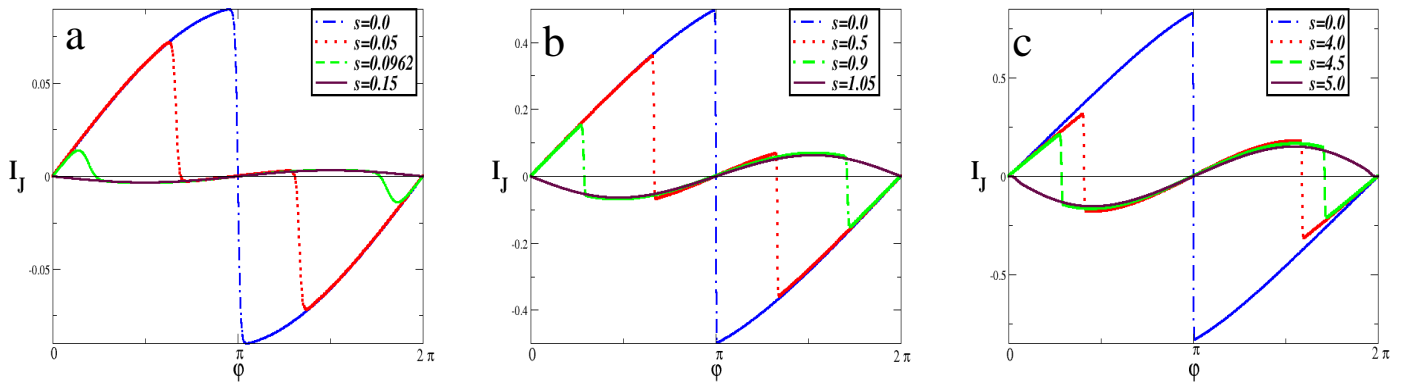}}
\caption{ The total Josephson current ($I_J$)as function of the
phase difference for increasing spin. The $\pi$ junction behavior
is clearly seen. (a) $\Gamma=0.2$. (b)$\Gamma=2$. (c)$\Gamma=10$.
The other parameters are: $\Delta=1.0, \beta=1000, \gamma=0, $ and
$\epsilon=0$. }
\label{fig:IJphi}
\end{figure}

\begin{figure}[!]
\vskip 0.5in
\protect\centerline{\epsfxsize=4.5in\epsfbox{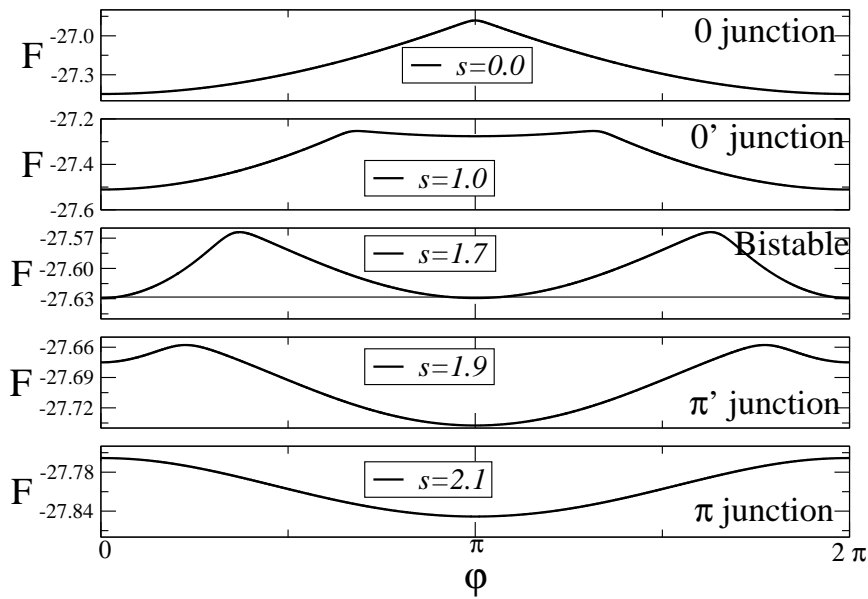}}
\caption{ The Free energy (F) as function of the phase difference
($\Phi$) for increasing spin from top to bottom. The $\pi$
junction behavior is clearly seen. The other parameters are:
$\Delta=1.0, \Gamma=4.0, \gamma=0, \beta=1000$ and $\epsilon=0$. }
\label{fig:Fphi}
\end{figure}

In Fig.~\ref{fig:IJphi}, we plot the Josephson
current($I_J$) as a function of the phase $\phi$, for different
values of the spin coupling $s=J S$. We clearly see a
$\pi$-junction type behavior as the magnitude of the spin coupling
is increased. One clear inference from Fig.~\ref{fig:IJphi} is
that the strength of the spin coupling required to engineer a
$\pi$ junction behavior increases with the interface transparency
$\Gamma$. We get another point of view of this $\pi$ shift in
Fig.~\ref{fig:Fphi}, where we plot the free energy $F$ as a
function of the phase $\phi$. We see that with increase in spin
coupling strength the transition from the $0$ to the $\pi$ phase
is clearly marked: the absolute minimum of $F$ shifts from
$\phi=0$ to $\phi=\pi$.

This figure also brings out other features, namely the $0'$ and
$\pi'$ phases. As is evident from Fig. 5, the labelling of the
respective junctions as $0, 0', \pi '$ and $\pi$ configurations,
follows from the respective stability of $\phi=0$ and $\phi=\pi$
configurations. For a $0(\pi)$ junction, only $\phi=0 (\phi=\pi)$
is a minimum of $F(\phi)$. For the other two cases, both $\phi=0,
\pi $ are local minima and depending on whether $\phi=0
(\phi=\pi)$ is the global minimum, one has a $0' (\pi ')$
junction.\cite{arovas} Of particular interest is the bistable
junction, in which both $\phi=0$ and $\phi=\pi$ are global minima;
note that this bi-stability is a necessary precondition for the
realization of a Josephson junction qubit.\cite{ioffe}

 The total Josephson current and the free energy are invariant
 with respect to the change of sign of the spin ($s \to -s$). As explained
in part B of the previous section, this is to be expected from the invariance of the system
under the spin up - spin down exchange. Technically, it can be seen
on Eqs.~(\ref{eq:IJmatsu})-(\ref{free_matsu}), using $\omega_{-(n+1)} = - \omega_n$.
Note that, as explained before, the Andreev bound states are not invariant under $s \to -s$,
and thus the Andreev bound states current and the continuum current are also not invariant.


\subsection{Mechanism of the $\pi$-shift}

\begin{SCfigure}
 \centering
\includegraphics[totalheight=10in,width=4.54in]{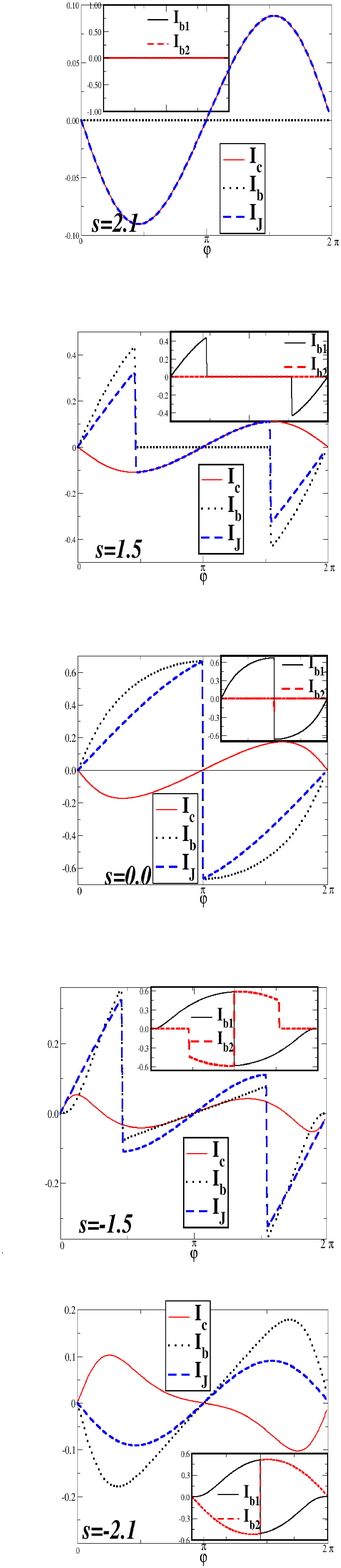}
 \caption{The Andreev bound state current ($I_b$), continuum
current ($I_{c}$) and total Josephson current ($I_{J}$) as
function of phase difference $\phi$. The individual Andreev bound
state contributions ($I_{b1},I_{b2}$) are plotted in the insets.
Central figure is for spin $s=0.0$. The two figures to the bottom
of the central figure depict the case of $s<0$,
 the first is for spin $s=-1.5$ and last one on
the bottom is for $s=-2.1$. On the top of the central figure the
two figures depict the case of $s > 0$. The
individual figures plotted are for spin $s=1.5$ and lastly for
$s=2.1$. The other parameters are: $\Delta=1.0, \Gamma=4.0,
\gamma=0.0, \beta=1000$ and $\epsilon=0$. }
\label{fig:IbIc}
\end{SCfigure}

The ability to distinguish, in the Josephson current, between the
contributions from each Andreev bound state and from the continuum
(see Eq.~(\ref{eq:IJ})) provides us with a simple picture for the
mechanism leading to the $\pi$ shift for large spin coupling. 
Note that the picture we obtain depends on the initial choice of spinors
(see the discussion at section III D); the choice we make here allows
us to get a very simple picture. 
In a few words, the effect of the spin coupling is to reduce or
suppress the Andreev bound states contribution, and to give more
importance to the continuum contribution, and this leads to the
$\pi$ shift. With more details, the effect of the spin coupling on
the bound states current can be understood from Eq.~(\ref{eq:IJ})
and Fig.~\ref{fig:Andreevlevelsphi}. For $s=0$, we see in
Fig.~\ref{fig:Andreevlevelsphi} that there is always one bound
state below the Fermi level, and the other one above. As the
contribution of a bound state to the Josephson current is
proportional to the occupation number $n_F(E_i)$ of this bound
state (Eq.~(\ref{eq:IJ})), this means that we have only one bound
state contributing to the current, and this contribution appears
to be much larger than the continuum contribution. With a large
positive spin coupling $s$, we see on
Fig.~\ref{fig:Andreevlevelsphi} that both Andreev bound states are
above the Fermi level, which means their contribution to the
Josephson current vanishes; while for a large negative spin
coupling $s$, we see that both bound states are below the Fermi
energy, which means they both contribute to the Josephson current,
and this reduces the total bound state contribution as the
respective contributions of the two bound states have opposite
signs. Note that the total Josephson current is independent of
the spin coupling sign, but that for large $s>0$ there is only the
contribution from the continuum, while for large $s<0$ there is a
combinations of the bound states and the continuum contributions.
This explanation is illustrated on Fig.~\ref{fig:IbIc}, where the
contributions of the bound states and of the continuum are plotted
for different values of $s$.

The origin of the continuum current -which is non-zero even at
 zero temperature- is due to the phase difference between
 the two superconductors, which breaks the symmetry between
the left and right-moving
quasi-particles.\cite{bagwell92,continuum}
 One can draw  an analogy with persistent currents flowing
in normal metal rings at zero
temperature. In normal metal rings the flux breaks the symmetry
between clockwise and anti-clockwise moving electrons inducing the
persistent current. At zero temperature all states below the Fermi
energy are filled, still then the persistent current is
non-zero.\cite{buttiker}

We also observe that the continuum current generally flows opposite
to the bound state current. This observation is in agreement
with that of other works.\cite{kamenev,kirch}

We finally add that one can also understand the fact that the full current is the same for $+s$
and $-s$ but with different contributions from bound states and continuum by
using electron hole symmetry. At first sight, electron hole symmetry only holds
when the dot level coincides with the superconducting chemical potential.
We first discuss this case, and then we address below the case where a gate voltage
shifts the dot level away from this location.   

Consider the case of negative coupling ($s<0$) in Fig.~\ref{fig:Andreevlevelsphi}a.
From the electron point of view, occupied states below the Fermi level have a continuum contribution
to the current and a bound state contribution. From the point of view of holes, which occupy
all states above the Fermi level, there is a continuum contribution to the current
of occupied hole states while the contribution of bound states above
the Fermi level diminishes with increasing $|s|$. For sufficiently large $|s|$, the
two bound states are below the Fermi level and the holes cease to have a bound state contribution.
On the opposite, for positive coupling in  Fig.~\ref{fig:Andreevlevelsphi}b, the role of electrons and
holes is reversed: from the electron point of view, the bound state
contribution is reduced --and eventually vanishes -- when increasing $s>0$; from the hole
point of view, holes occupying both the continuum and the Andreev bound states
above the Fermi level. We thus see that for positive (negative) $s$, the role
of electrons and holes is reversed, and the electron hole symmetry can explain
why physically observable quantities are invariant under the substitution
$s\to -s$.

Next consider the case where the dot level does not coincide with the 
chemical potential ($\epsilon\neq 0$). In any normal metal devices, this indeed breaks 
electron hole symmetry, nevertheless we wish to point out that because 
we are dealing with a superconducting system the above argumentation still 
holds. When dealing with an arbitrary superconductor-normal metal- superconductor
(SNS) junction, Andreev bound states are understood from the conversion of electrons
into holes and vice versa at the superconducting junction. An alternative picture 
is to say that two electrons, one above the chemical potential with energy $E$, 
and one below with energy $-E$ are transfered through the normal region from 
one superconductor to the other. 
For our quantum dot setup, calculations of the 
Josephson current could for instance be performed using T-matrix formalism
(to all orders for an exact result), 
where information about energies appears only in the form of differences of 
energy levels ($E-\epsilon$ for the positive energy electron and $E+\epsilon$
for the negative energy electron). 
These two energy differences are invariant under the change $\epsilon\to -\epsilon$.
A dot with say, $\epsilon>0$ will thus have the same
Andreev bound states spectrum  and Josephson current, as a dot with $\epsilon<0$.      
This is indeed explicit in Eq. (\ref{eq:detM}), which is 
invariant under the transformation $\epsilon\to -\epsilon$.
In the presence of the impurity spin the bare dot level is split by 
a Zeeman like coupling (the exchange term), nevertheless the above statement 
($\epsilon\to -\epsilon$ symmetry in the energy differences) still applies: 
the Andreev spectrum is reversed between $+s$ and $-s$, but the Josephson 
current is shown to be the same because it can be computed from the point of 
view of electrons (say, for $+s$) or, alternatively, of holes (for $-s$).    

In conclusion, because the Andreev bound state is the same for  positive and negative 
$\epsilon$, the symmetry $+s\to -s$ also holds for physically observable 
quantities such as the current. The argument for $\epsilon=0$
describing the contribution of electrons and holes populations on the continuum and 
discrete levels is unaffected by the modification $\epsilon\neq 0$.
This fact is demonstrated in Fig. \ref{fig:IJepsilon}c, 
where the bound state spectrum is plotted as a function of flux 
for several values of $\epsilon$, for positive and negative $s$.

\subsection{Controlling the $\Pi$ shift}

\begin{figure}[!]
\protect\centerline{\epsfxsize=8.0in\epsfbox{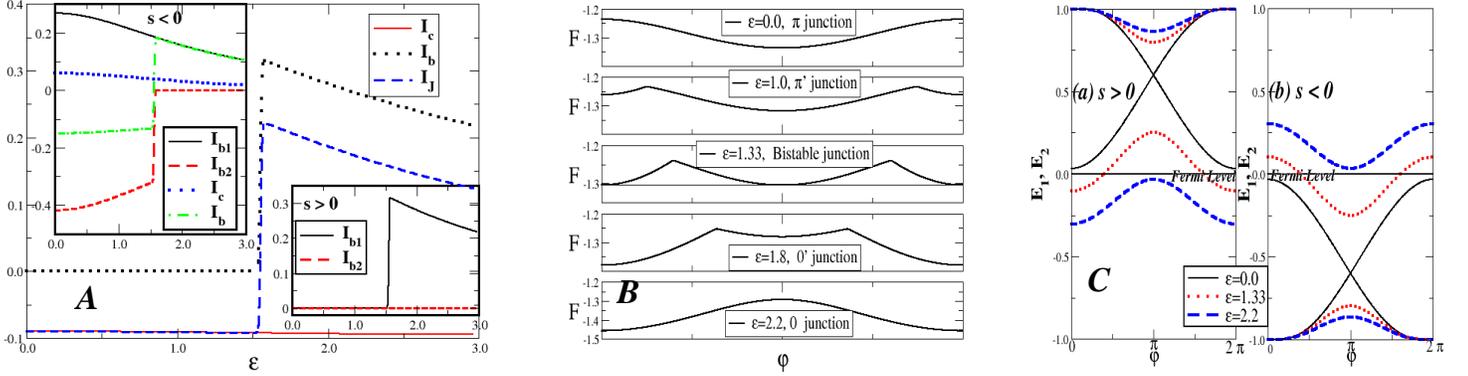}}
\caption{ (A) The Andreev bound state current ($I_b$, dotted
line), the continuum contribution ($I_{c}$, solid line) and the
total Josephson current ($I_{J}$, dashed line) as function of dot
level $\epsilon$, for positive coupling. The $\pi$ junction
behavior is clearly seen as dot level is varied, in the insets the
individual bound state contributions($I_{b1}, I_{b2}$) are plotted
for $s < 0$ and positive $s> 0$  (for $s<0$ the continuum and total bound
state currents are also plotted). The total Josephson current
($I_J$) is identical for $s < 0$ and $s > 0$. The
other parameters are: $\Delta=1.0, s=2.1, \phi=\pi/2, \Gamma=4.0,
\beta=1000$ and $\gamma=0$. (B) The Free energy as function of the
phase difference ($\phi$) for increasing dot energies, and (C)
Andreev levels for both negative as well as positive coupling. The
other parameters for B and C are: $\Delta=1.0, s=2.1, \Gamma=4.0,
\beta=1000$ and $\gamma=0$. }
\label{fig:IJepsilon}
\end{figure}

\begin{figure}[!]
\protect\centerline{\epsfxsize=7.5in\epsfbox{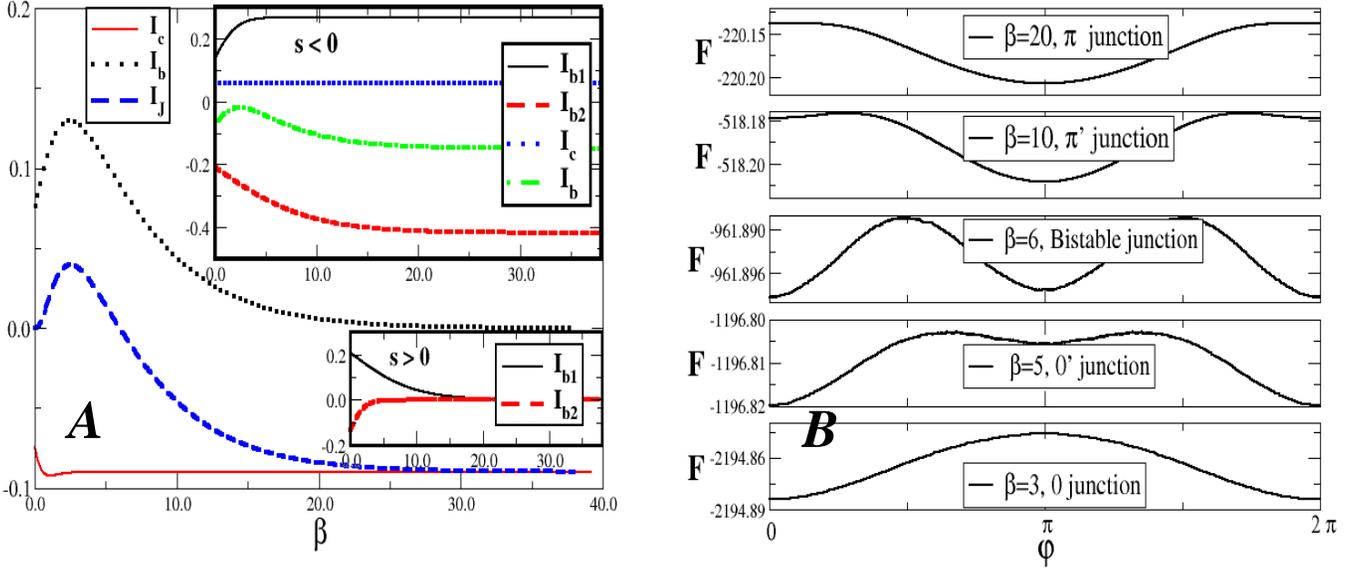}}
\caption{ (A) Andreev bound state current ($I_b$), the continuum
contribution ($I_c$) and the total Josephson current ($I_J$) as
function of inverse temperature $\beta$ for positive coupling. The
$\pi$ junction behavior is clearly seen as inverse temperature
$\beta$ is varied. In the insets the individual bound state
currents for the case of $s > 0$ and $s < 0$ are plotted. For $s <
0$, the continuum along with the total bound state current is also
plotted. The total Josephson current $I_J$ is identical for
$s > 0$ and $s < 0$. The other parameters are: $\Delta=1.0, s=2.1,
\epsilon=0.0, \gamma=0.0, \phi=\pi/2$ and $\Gamma=4.0$.(B)The Free
energy as function of the phase difference at different
temperatures. The junction is bistable at the crossover between
$0-\pi$ junction behaviors. The other parameters are: $\Delta=1.0,
s=2.1, \epsilon=0.0, \gamma=0.0,$ and $\Gamma=4.0$. }
\label{fig:IJtemp}
\end{figure}

A remarkable feature of our system is that the $\pi$ shift
behavior can be controlled and reversed using the different
parameters of the system. This is important for potential
experimental implementations, as  some of these parameters can be
accessed relatively easily (one can for example move the dot level
by using a gate voltage~\cite{park}), while the spin coupling is a
fixed quantity which depends in the molecule used. Our results
show that, when the spin coupling is large enough to have a $\pi$
junction, a change in any of the parameter of the system (dot
level $\epsilon$, coupling to the leads $\Gamma$, asymmetry of
this coupling $\gamma$, and even the temperature) makes it
possible to have the system behave as a standard $0$ junction
(going through any intermediate situation between $\pi$ and $0$
junction). Schematically, the mechanism for this can be understood
along the same lines as the explanation given above for the $\pi$
shift: starting from a $\pi$ junction situation, where both
Andreev levels are (for example) above the Fermi energy and  thus
do not contribute to the Josephson current, changing a parameter
of the system can move the Andreev level positions, and as soon as
one of the Andreev level goes below the Fermi energy, it gives an
important bound state contribution which brings the system back to
a $0$ junction behavior. This is illustrated on
Fig.~\ref{fig:IJepsilon}, where the dependence of the currents
(panel A), of the free energy (panel B) and of the Andreev levels
(panel C) as a function of the dot level position $\epsilon$.
Similar plots are obtained when looking at the $\Gamma$ or
$\gamma$ dependence (not shown).

The picture is a bit different when the temperature is changed, as
there the Andreev levels do not move, but the Fermi functions
become broader as temperature is increased, leading to a partial
revival of the bound state current. This is shown on
Fig.~\ref{fig:IJtemp}, where the dependence of the currents (panel
A) and of the free energy (panel B) is shown as a function of
$\beta= 1/(k_B T)$. Starting from low temperature (high $\beta$),
with a $\pi$ junction behavior (the total current $I_J$ is $<0$,
and the free energy has its minimum at $\phi=\pi$), we see that
when the temperature increases ($\beta$ decreases), the total
current becomes positive, and the minimum of the free energy
shifts from $\phi=\pi$ to $\phi=0$.

\section{Discussion}

We have studied in the previous sections the behavior of the
Josephson current as a function of the spin coupling strength, and
found that a $\pi$ junction behavior appears when this coupling is
large enough. In view of an experimental realization, one must ask
if the actual value of the spin coupling obtained with a given
molecular magnet is large enough to observe this $\pi$ junction
behavior. While a precise estimate, for a real molecule, of the
magnetic coupling energy between the electronic spin and the
molecular spin is beyond the scope of this paper,
 we can get a gross estimate by calculating the interaction energy
of two magnetic dipoles at a distance typical of the molecular
distance involved in our problem. Taking a spin $S=10$ for the
molecule (as in Mn12ac), and a distance $\sim5$ \AA,  we find a
interaction energy $\sim0.1 meV$, which is of the same order as
the superconducting gap. This estimate shows that the $\pi$
junction regime due to spin coupling may be reached
experimentally.

Let us now discuss some potential applications of our results.
The system could beused as a Josephson current switch. Looking at the panel A of
Fig.~\ref{fig:IJepsilon}, we see that there is an abrupt change of
the current sign as $\epsilon$ goes through a specific value
depending on the other parameters (it is $\epsilon\simeq 1.5$ on
the figure), while the current does not change much elsewhere.  As
$\epsilon$ should be experimentally accessible using a gate
voltage, a Josephson current switch could be implemented.
Moreover, this implementation should be easier than in systems
where the Josephson current changes sign several times as a
parameter is varied.

A more ambitious application would be to engineer a qubit
with the system we describe in this work. Indeed, we have shown
that, when varying some parameters, it is possible that the system
behaves as a bistable junction (see for example panel B of
Fig.~\ref{fig:IJepsilon}), where the system has a degenerate
ground state. This feature can be effectively exploited to fashion
a qubit system\cite{illichev}, where the junction itself can be in
a superposition\cite{ioffe} of the two ground states at either a
phase difference of $0$ or $\pi$. In contrast to the
superconducting persistent current qubit\cite{mooij}, it is here
the two phase states of the Josephson junction which provide the
two states of the qubit. These qubits are therefore called
superconducting phase qubits as in Ref.[\onlinecite{ioffe}].
Similar to that in Ref.[\onlinecite{feigelman}], the coherent Rabi
oscillations in our system could in principle be observed by a
measurement of the phase sensitive sub-gap Andreev conductance
across a high resistance tunnel contact between the qubit and a
dirty metal wire\cite{amin}.

\section{Conclusion}
To conclude, we have studied in this work the properties of the
Josephson current between two superconductors through a single
molecular magnet, which we modeled as a quantum dot plus a large
frozen spin. We have shown that the coupling between the
electronic spin on the dot and the molecular spin lead the system
to behave as a $\pi$ junction. We have given a simple mechanism
explaining this $\pi$ junction behavior, in terms ob bound state current and continuum current.

We have shown moreover that the other parameters of the system
give a precise control of this $\pi$ junction, allowing for
example to reverse the $\pi$ shift and to bring the system to the
normal $0$ junction state, or to an intermediate bistable state.
This control of the $\pi$ shift can lead to useful applications,
like a Josephson current switch, or could even be used to engineer
a phase qubit.

 Possible topics of future study in such systems may
include incorporating the dynamical nature of molecular
spin\cite{balatsky} and quantum tunneling of the
magnetization\cite{gatteschi}, when the anisotropy barrier is not
much larger than all the other energies of the problem.
Also interesting would be to include electron-electron/electron-phonon
interactions in the present work, in order to study the combination of
the effects of the molecular spin and of the Coulomb charging energy,
or/and the effect of electron-phonon interactions.

Centre de Physique Th\'eorique is UMR 6207 du CNRS, associated with
Universite de la M\'editerann\'ee, Universit\'e de Provence, and
Universit\'e de Toulon.

\acknowledgments The authors would like to acknowledge Dr. Eric
Soccorsi for valuable mathematical comments.

\end{document}